\documentclass[%
prb,twocolumn,superscriptaddress,
nofootinbib,
 amsmath,amssymb,
 aps,
]{revtex4-2}

\usepackage{physics}
\usepackage{graphicx}% Include figure files
\usepackage{dcolumn}% Align table columns on decimal point
\usepackage{bm}% bold math
\usepackage{hyperref}% add hypertext capabilities
\usepackage[USenglish]{babel}
\usepackage{comment}
\usepackage{float}
\usepackage{gensymb}
\usepackage{color}

\usepackage{subcaption}
\usepackage{mwe}
\begin{document}

\preprint{APS/123-QED}

\title{Out-of-plane magnetic anisotropy in bulk ilmenite $\text{CoTiO}_3$}

\author{M. Arruabarrena}%
\affiliation{Centro de F\'\i sica de Materiales - Materials Physics Center (CFM-MPC), 20018 Donostia, Spain}
\author{A. Leonardo}
\affiliation{Department of Applied Physics II, University of the Basque Country UPV/ EHU, Bilbao, Spain}
\affiliation{Donostia International Physics Center (DIPC), 20018 Donostia, Spain}
\author{M. Rodriguez-Vega}
\affiliation{Theoretical Division, Los Alamos National Laboratory, Los Alamos, New Mexico 87545, USA \looseness=-1}

\author{Gregory A. Fiete}
\affiliation{Department of Physics, Northeastern University, Boston, MA 02115, USA}
\affiliation{Department of Physics, Massachusetts Institute of Technology, Cambridge, MA 02139, USA}

\author{A. Ayuela}
\affiliation{Centro de F\'\i sica de Materiales - Materials Physics Center (CFM-MPC), 20018 Donostia, Spain}
\affiliation{Donostia International Physics Center (DIPC), 20018 Donostia, Spain}

\date{\today}% It is always \today, today,

\begin{abstract}
Structural, electronic and magnetic properties of bulk ilmenite CoTiO$_3$ are analyzed in the framework of density functional theory (DFT), using the generalized gradient approximation (GGA) and Hubbard-corrected approaches. 
We find that the G-type antiferromagnetic (G-AFM) structure, which consists of antiferromagnetically coupled ferromagnetic $ab$ planes, is the ground-state of the system, in agreement with experiments. Furthermore, cobalt titanates present two critical temperatures related to the breaking of the inter- and intra-layer magnetic ordering. This would result in the individual planes remaining ferromagnetic even at temperatures above the N\'{e}el temperature.
When spin-orbit coupling is included in our calculations, we find an out-of-plane magnetic anisotropy, which can be converted to an in-plane anisotropy with a small doping of electrons corresponding to about 2.5\% Ti substitution for Co, consistent with experimental expectations.  We thus present a disorder-dependent study of the magnetic anisotropy in bulk $\text{CoTiO}_3$, which will determine its magnon properties, including topological aspects.
\end{abstract}

\maketitle

\section{Introduction}

Titanate materials ATiO$_3$ (with A = a rare earth or transition metal element) have a wide variety of crystal structures, which result in numerous intriguing physical phenomena such as ferroelectricity, magnetism, multiferroicity, and piezoelectricity\cite{Perovskites}. In particular, cobalt titanate, CoTiO$_3$, has a broad variety of electronic based industrial applications including  catalysis \cite{Catalyst_1}, as a high-$\kappa$  dielectric\cite{High-k} (where $\kappa$ is the dielectric constant), and as a gas sensor \cite{Applications_SIEMONS}. In addition, CoTiO$_3$ has been reported to exhibit Dirac magnons\cite{DiracMagnons} and a magnetodielectric effect\cite{DUBROVIN2020157633}. 
Despite the growing interest in the electronic and magnetic properties of cobalt titanate, to the best of our knowledge, first-principles theoretical studies of its magnetic properties are absent in the literature.  

Magnetic properties of CoTiO$_3$ ilmenites are ascribed to cobalt atoms in the form of Co$^{2+}$ ions distributed in layers, structurally in a $C_{3v}$ symmetry given by the neighbouring oxygen atoms.
Magnetic susceptibility studies indicate that cobalt magnetic moments are antiferromagnetically coupled between layers while they are ferromagnetically coupled within layers\cite{Newnham,WATANABE1980}.
Neutron diffraction experiments assign in-plane magnetic moments to cobalt atoms, a fact that lowers the symmetry around cobalt atoms\cite{Newnham, DiracMagnons}.
However, these studies find that magnetic excitations recover the $C_{3v}$ symmetry around Co$^{2+}$ ions.

To reconcile the two pictures, these experimental works have assumed models that include in-plane structural domains given by staggered trigonal distortions and oxygen twin planes. Neutron scattering averages over the domains and allows one to recover the $C_{3v}$ symmetry found in magnetic excitations. Using first principles calculations, the phonon vibrational properties were studied to explain Raman observations\cite{DUBROVIN2020157633}. Therefore, to complement these lattice dynamics results, and clarify the validity of the assumptions made to explain neutron scattering data, there is a need to study the magnetic properties of CoTiO$_3$ ilmenites in a single perfect crystal.

In this paper, a systematic DFT-based first-principles analysis of the structural, magnetic, and electronic properties of CoTiO$_3$ is performed. In the framework of the Hubbard-corrected GGA (GGA+U)\cite{DFT+U-Tolba}, we calculated the lattice parameters and band structure of CoTiO$_3$. 
We found that the G-type antiferromagnetic structure reported in the experiments \cite{Newnham,DiracMagnons,WATANABE1980} is the ground-state of the system. Two critical temperatures are observed, resulting in a temperature region above the N\'{e}el temperature where the system would still present ferromagnetism within the individual layers. 

We also calculated the magnetic anisotropy of the system, which favors out-of-plane magnetization, a finding that seems to be at odds with previous experimental findings\cite{Newnham,DiracMagnons,WATANABE1980}. However, we analyze the variation of the magnetocrystalline anisotropy energy (MAE) with respect to the number of electrons in the unit cell, and propose that the experimental in-plane magnetization could be a result of doping in the system. We compute the low-doping level that would produce the change to in-plane magnetization.

This paper is organized as follows.  In Sec.\ref{sec:theory_detail} we provide the details of our theoretical analysis, including the computational methodology for the chemical and magnetic structure determination.  In Sec.\ref{sec:results} we present the main results of our theoretical study, including the dependence of the structural, electronic, and magnetic properties on the correlations on the Co and Ti atoms.  We also discuss the magnetic anisotropy and the influence of the electron density on the magnetic properties.  Finally, in Sec.\ref{sec:conclusions}, we present the main conclusions of this work. 

\begin{figure*}
\makebox[\textwidth]{
   \centering
    \includegraphics[scale=0.085]{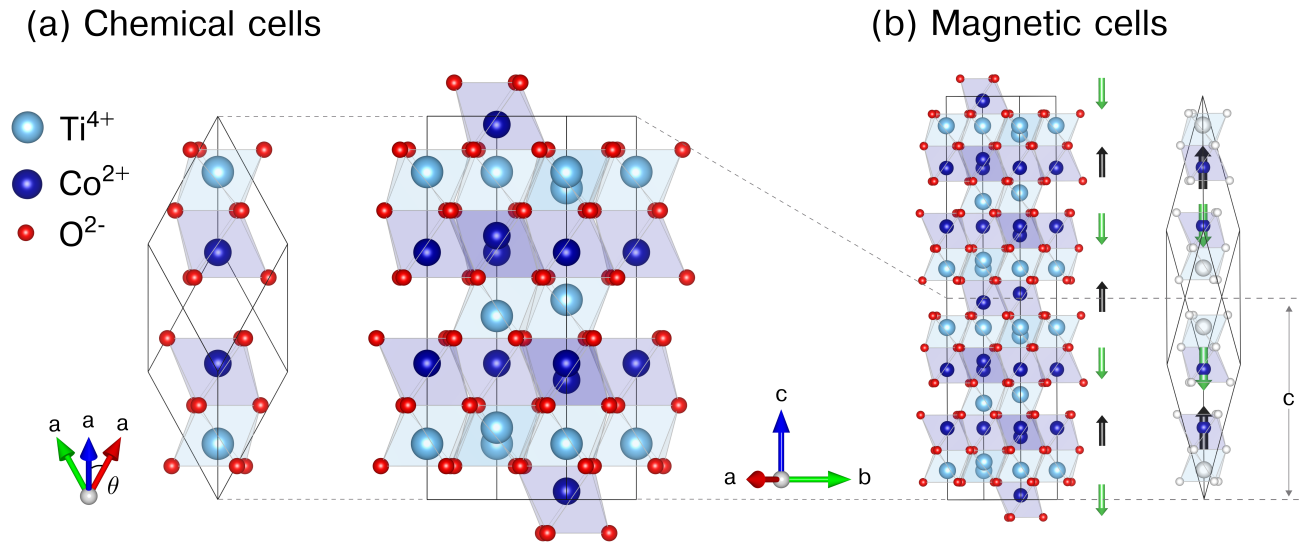}}
    \caption{Chemical and magnetic unit cells of bulk CoTiO$_3$. (a) The rhombohedral and hexagonal cells that reproduce the crystalline chemical periodicity. (b) Conventional and primitive magnetic cells exemplified using the experimentally found  ``G-AFM" magnetic configuration. Ferromagnetic hexagonal $ab$ planes of cobalt are antiferromagnetically coupled in the (doubled) hexagonal $c$-axis. This figure was prepared using the  VESTA software \cite{VESTA}.}
    \label{fig:structure}
\end{figure*} 

\section{Theoretical Approach}
\label{sec:theory_detail}
\subsection{Computational methodology}

Electronic delocalization within DFT may lead to an incorrect description of the magnetic properties.  In particular, for systems with localized electrons such as $d$-electrons of transition metals acting as dopants in semiconductors or constituting a component of transition metal oxides, Coulomb interaction effects may lead to qualitatively different results \cite{Raebiger,Electron-overdelocalization-doi:10.1063/1.4922693,Raebiger,aguilera2019magnetism}. The DFT+U method is one approach that aims to correct the tendency of DFT towards itineracy by explicitly correcting the Coulomb interaction with a Hubbard-like interaction for a subset of states in the system\cite{DFT+U-Tolba}. By including the on-site Coulomb interaction $U$ and exchange interaction $J$ terms, the non-integer or double occupation of these states is penalized, thus localizing them in the atomic sites. 

Our DFT calculations were performed using the Vienna Ab-initio Software Package (VASP)\cite{VASP1, VASP2} using the projector augmented wave method (PAW). We employed the GGA for exchange using  Perdew--Burke--Ernzerhof (PBE) approach. Extra electron-electron Coulomb interactions are taken into account with the GGA+U approach implemented in the code. We employed the simplified (rotationally invariant) approach by Dudarev {\it et al.}\cite{Dudarev-PhysRevB.57.1505}, which includes the $U$ and $J$ terms as an effective $U_{\text{eff}} = U - J$ parameter. For brevity, and unless stated otherwise, we refer to this $U_{\text{eff}}$ parameter as $U$ for the rest of the paper. 
The electrons Co(3$p$, 3$d$, 4$s$), Ti(3$p$, 3$d$, 4$s$) and O(2$s$, 2$p$) were treated as valence states.  Tests using all-electron calculations were conducted to check that the number of valence electrons per element were properly considered, as described in Appendix A.

For most of the calculations presented in this paper, the total energy of the system was converged with respect to the plane-wave cutoff energy and reciprocal space samplings. The convergence criterion was less than $1$ meV/atom, and we found that a plane-wave cutoff of $800$ eV, and a $\Gamma$-centered 8x8x8 Monkhorst-Pack $k$-point mesh to yield results within the stated precision. In the spin-orbit calculations, where the energy differences are on the order of $10^{-1}$ meV, additional convergence tests for the magnetocrystalline anisotropy energies (MAE) with respect to the reciprocal space sampling were performed to ensure numerically precise results (see Appendix E).

\subsection{Chemical and magnetic structures}

We first discuss the difference between the chemical and magnetic structures of bulk cobalt titanate. The compound CoTiO$_3$ is reported to have an ilmenite crystal structure with trigonal space group R3$^-$, which consists of alternating layers of corner sharing CoO$_6$ and TiO$_6$ octahedra, stacked along the $c$-axis in the hexagonal setting, as shown in Fig. \ref{fig:structure}(a)\cite{Newnham,DiracMagnons}. The experimental lattice parameters are $a$ = 5.48 $\text{\AA}$ and $\theta=55 \degree$ in the rhombohedral setting, and $a$ = $b$ = 5.06 $\text{\AA}$ and $c$ = 13.91 $\text{\AA}$ in the hexagonal setting. The  Co, Ti and O atoms are located at the Wyckoff positions (0,0, 0.355), (0, 0, 0.146) and (0.316, 0.021, 0.246), respectively \cite{DUBROVIN2020157633}. 

The CoTiO$_3$ magnetic configuration is reported as ``G-type" antiferromagnetic ordering below the N\'{e}el temperature of 38 K \cite{Newnham,DiracMagnons,DUBROVIN2020157633,WATANABE1980,SCHOOFS}. This configuration consists of ferromagnetically coupled hexagonal $ab$-planes, antiferromagnetically coupled along the $c$-axis, as shown in Fig.\ref{fig:structure}b. 
It should be noted that in order to reproduce the periodicity of this magnetic cell in the spin-polarized formalism implemented in the {\it ab initio} codes, building a cell larger than the chemical rhombohedral or hexagonal cells is needed. 

Although the $c$-axis doubled hexagonal cell, which consists of 60 atoms, is a straightforward candidate, there is a primitive magnetic cell of just 20 atoms that still satisfies this periodicity\cite{Elliot}. It can be defined by means of the transformation 
\begin{gather}
    \begin{bmatrix} \textbf{M}_{1} \\ \textbf{M}_{2} \\ \textbf{M}_{3} \end{bmatrix}
    = \frac{1}{3}
    \begin{bmatrix} 1 & 2 & 2 \\ -2 & -1 & 2 \\ 1 & -1 & 2 \end{bmatrix} 
    \begin{bmatrix}
    \textbf{a} \\ \textbf{b} \\ \textbf{c}
    \end{bmatrix}, 
\end{gather} 
where \textbf{a}, \textbf{b} and \textbf{c} are the hexagonal lattice vectors. We refer to this cell as the primitive magnetic cell, and unless stated otherwise, all the calculations in this paper are performed in this configuration.

\section{Results and discussion} 
\label{sec:results}

\subsection{Structural properties}

Using the primitive magnetic cell in the G-AFM configuration, the lattice parameters, cell volume, and atomic positions were fully relaxed for a range of different titanium and cobalt $U$ values. The stability of the structure was confirmed by additional phonon calculations that can be found in Appendix B. 
In Fig.\ref{fig:str}(a,b), the $c$ hexagonal lattice parameter and the cell volume are plotted against the $U$ parameter. Panels (c) and (d) display the values of the Co-Co and Co-Ti distances. The lattice parameter \textit{c} and volume \textit{V} are presented in the hexagonal setting in order to facilitate the interpretation.

Our results indicate that the GGA+U approach consistently overestimates the experimental lattice parameters, which is manifested in the volume expansion of the unit-cell with increasing $U$ parameters. This expansion ranges from 1.2\% of the experimental cell in the bare GGA case, to a 6.44\% value for the GGA+U case with (U$_{\text{Ti}}$=6, U$_{\text{Co}}$=5). This trend is in agreement with other investigations performed for TiO$_2$ in the framework of the DFT+U theory\cite{DFT+U}.  

The volume increase is linked to the expansion of the hexagonal $c$-axis, which is in turn closely related to the Co-Co and Co-Ti interatomic distances. However, the Co-Co distance decreases for small $U_{\text{Co}}$, before stabilizing, while the Co-Ti distance increases for all $U_{\text{Co}}$. In both cases, the effect of the titanium parameter $U_{\text{Ti}}$ is to decrease the distance with decreasing $U_{\text{Ti}}$. Note that the trends of these distances with $U$ are opposite in cobalt and titanium, as they are respectively above and below half-filling of the $d$-shell. The Ti-O bonds are the key ones determining the expansion in volume.

\begin{figure}
    \centering
    \makebox[0.5\textwidth]{
    \includegraphics[scale=0.25]{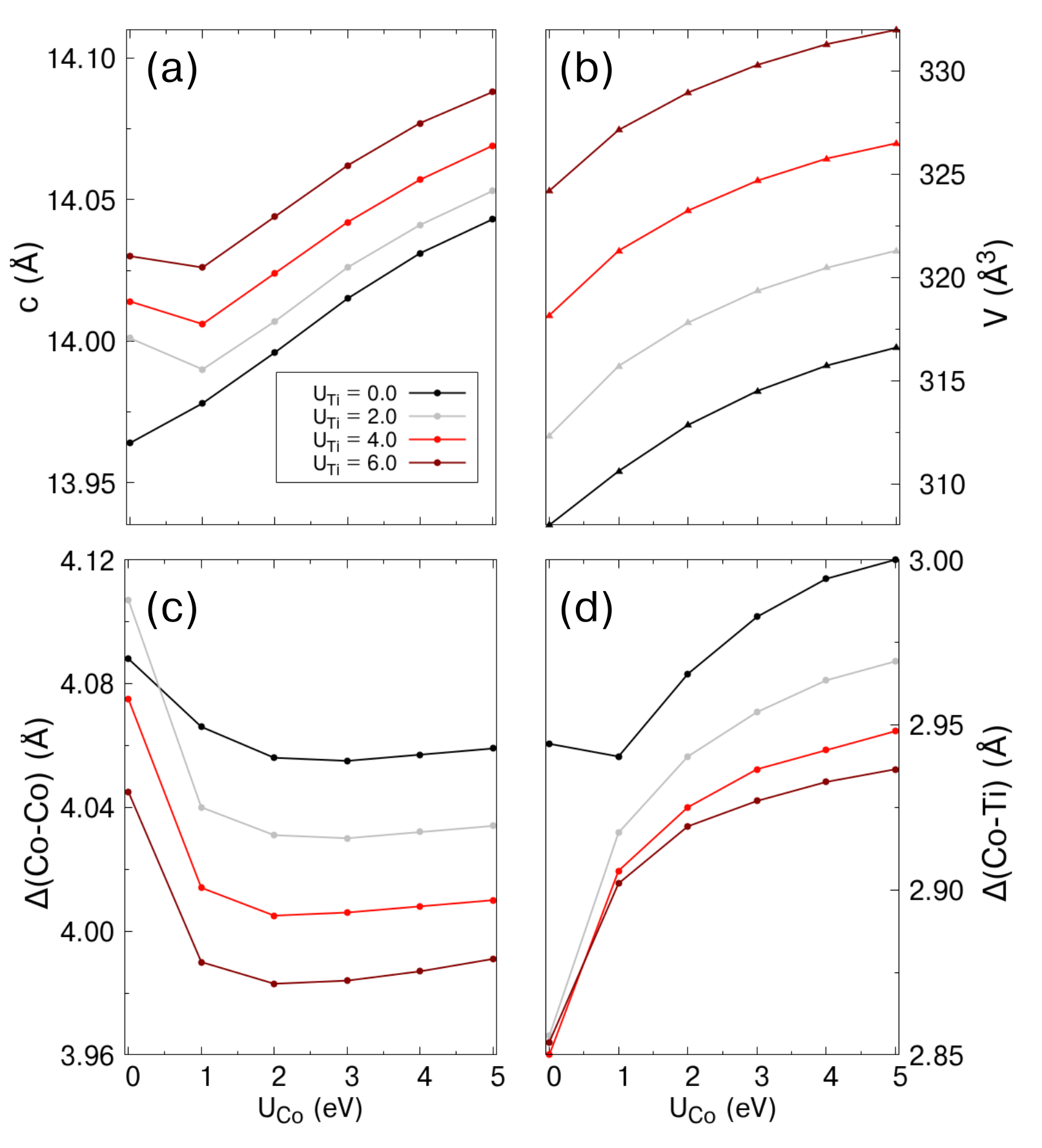}}
    \caption{Change of the structural parameters with respect to the chosen $U$ Coulomb-exchange values: (a) hexagonal lattice parameter $c$, (b) cell volume, (c) Co-Co distance, and (d) Co-Ti distances. }
    \label{fig:str}
\end{figure}

%%%%%%%%%%%%%%%%%%%%%%%%%%% %%%%%%%%%%%%%%%%%%%%%%%%%%% %%%%%%%%%%%%%%%%%%%%%%%%%%%  
%%%%%%%%%%%%%%%%%%%%%%%%%%%     ELECTRONIC PROPERTIES   %%%%%%%%%%%%%%%%%%%%%%%%%%% 
%%%%%%%%%%%%%%%%%%%%%%%%%%% %%%%%%%%%%%%%%%%%%%%%%%%%%% %%%%%%%%%%%%%%%%%%%%%%%%%%% 

\begin{figure}
\centering
\makebox[0.475\textwidth]{%
\includegraphics[scale=0.45]{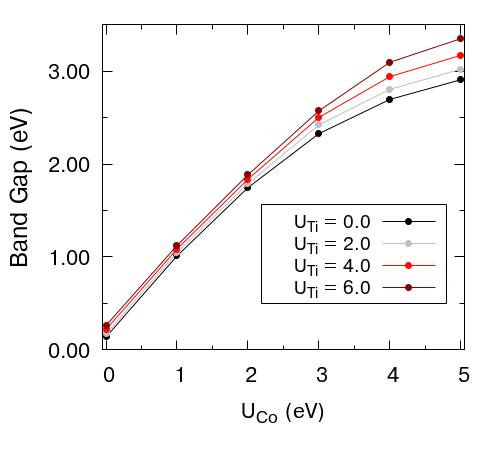}}
\caption{\label{fig:bandgap}Band gap as a function of cobalt and titanium $U$ parameters $U_{Co}$ and $U_{Ti}$, given in eV.}
\end{figure}  

\subsection{Electronic properties}

\begin{figure*}[]
    \centering
    \includegraphics[scale=0.32]{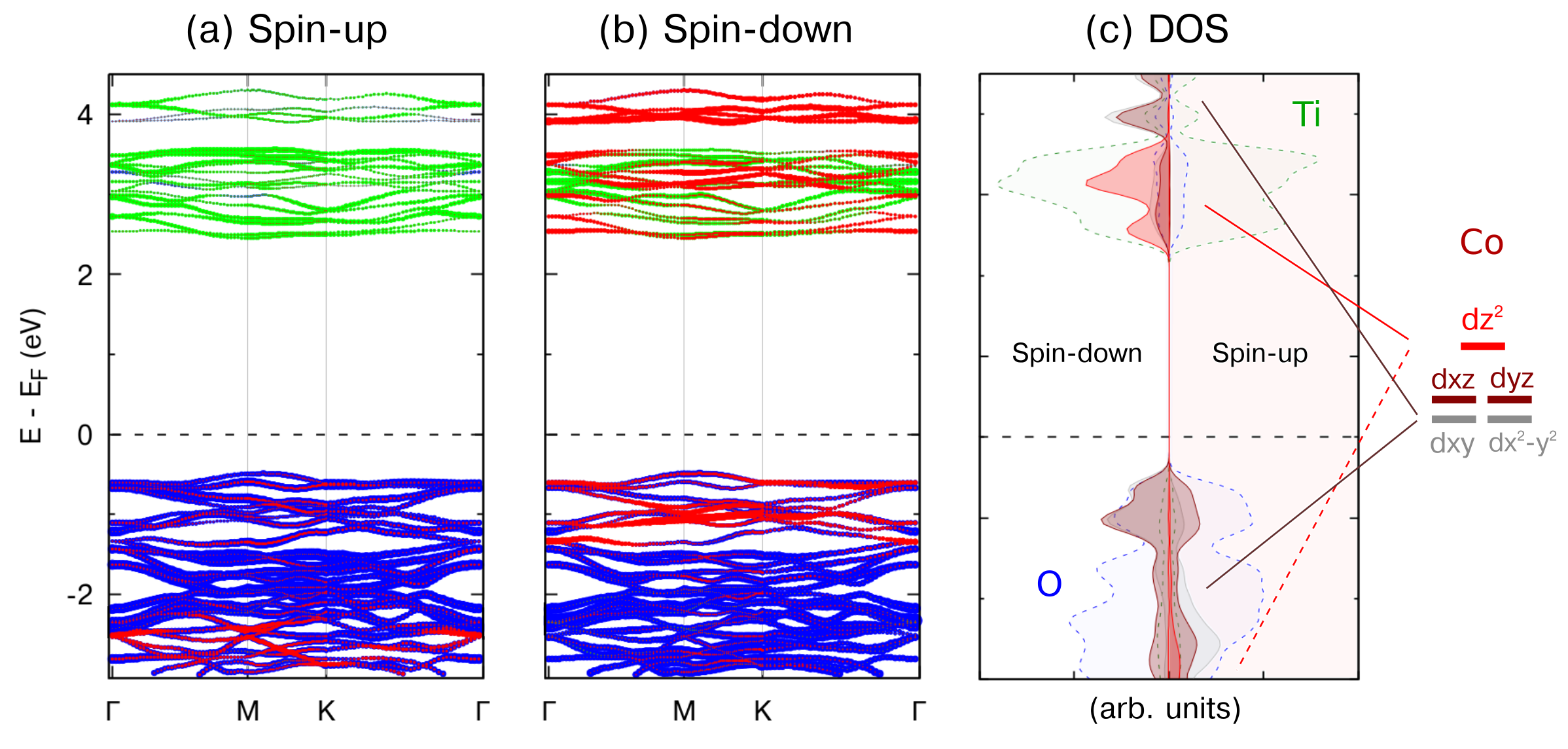}
    \caption{Element- and orbital-projected band-structure of cobalt titanate. The size of the marks represents the fractional contribution of each element following a color code. Titanium($d$) and oxygen($s+p$) are shown in green and blue, respectively. In red, the contribution of the $d$-orbitals of a cobalt atom in a spin-up G-AFM layer is shown. 
    Panel (c) shows the orbital-projected density of states (DOS) of a single cobalt atom in conjunction to the titanium and oxygen densities, as well as a sketch of the splitting of cobalt $d$-bands. $U$ parameters $U_{\text{Co}}$ = $U_{\text{Ti}}$ = 4.0 eV were employed in this calculation. }
    \label{fig:bands_evolution}
\end{figure*}

In order to investigate the electronic properties of the material, the band structure of CoTiO$_3$ was calculated. As in previous sections, we focused on the effect of the parameters $U_{\text{Co}}$ and $U_{\text{Ti}}$ on the electronic states of the system. Figure \ref{fig:bandgap} shows the change in the band gap with respect to these parameters. Figure \ref{fig:bands_evolution} displays the element-projected band structures in the GGA+U approach. The density of states (DOS) is also shown in the GGA+U case, as well as a sketch of the  valence and conduction bands of cobalt titanate. For comparison the band structures using GGA are included in Appendix C.

In Fig.\ref{fig:bandgap}, we observe how the cobalt $U_{\text{Co}}$ parameter increases the band gap, yielding values that range from an almost metallic state in the GGA case, to an insulating behavior with band gaps of the order of 3 eV. 
The effect of the titanium parameter $U_{\text{Ti}}$ is, in contrast, much more moderate, making appreciable differences only in the cases with a $U_{\text{Co}}$ larger than 3.0 eV. For $U$ values in the range $\simeq$ 3-4 eV, the bandgap takes values slightly below 3 eV. 

 In the GGA approach, 
the band gap of the system is given by cobalt $d$-bands close to the Fermi energy, with oxygen and titanium bands occupying the valence and conduction states, respectively. Hybridization seems negligible in the highest occupied valence band and lowest unoccupied conduction bands, and cobalt can be thought of as a recombination center,
as shown in Appendix C.  As the $U$ parameters increase, these cobalt bands around the Fermi energy become more localized, enhancing the energetic separation between them and increasing the insulating behavior of the system, as illustrated in Fig.\ref{fig:bandgap}.

In the GGA+U approach, shown in Fig. \ref{fig:bands_evolution}, these features are clearly observed. Cobalt titanate has a bandgap of 2.9 eV, and cobalt bands are integrated into the bulk continuum, mixing with oxygen and titanium in the valence and conduction bands, respectively. It should be noted that this mixing does not occur in a fully symmetric way, as the top of the valence band is characterized by Co-O states, while the bottom of the conduction band is mostly titanium-based. This explains the effect of $U_{\text{Ti}}$ in the electronic structure, which becomes notable when the band gap is given by the titanium bands in which this parameter acts.  
This finding would also have implications in the optical processes, as one would expect electrons to localize in the Ti-O layers, with holes concentrating in the Co-O layers, potentially leading to interesting excitonic behavior between the hexagonal $ab$-planes. 

%Comment DOS 
In the right panel of Fig. \ref{fig:bands_evolution}, we plot the projected density of states of a single cobalt atom along the densities of titanium and oxygen. We find how the out-of-plane orbital $d_{\text{z}^2}$ concentrates in the conduction band, while the in-plane $d_{\text{x}^2-\text{y}^2}$ and $d_{\text{xy}}$ orbitals are hybridized with the $d_{\text{xz}}$ and $d_{\text{yz}}$ orbitals over a wide energy range. 
This spatial distribution of the electronic states gives a picture of the bonding in the crystal, with the cobalt $d_{\text{z}^2}$ orbitals participating in the Ti-Co interaction, and the rest of the cobalt $d$-orbitals hybridizing with oxygen.

%%%%%%%%%%%%%%%%%%%%%%%%%%% %%%%%%%%%%%%%%%%%%%%%%%%%%% %%%%%%%%%%%%%%%%%%%%%%%%%%%  
%%%%%%%%%%%%%%%%%%%%%%%%%%%      MAGNETIC PROPERTIES    %%%%%%%%%%%%%%%%%%%%%%%%%%% 
%%%%%%%%%%%%%%%%%%%%%%%%%%% %%%%%%%%%%%%%%%%%%%%%%%%%%% %%%%%%%%%%%%%%%%%%%%%%%%%%% 

\begin{figure*} 
\centering
\makebox[0.5\textwidth]{%
\includegraphics[scale=0.15]{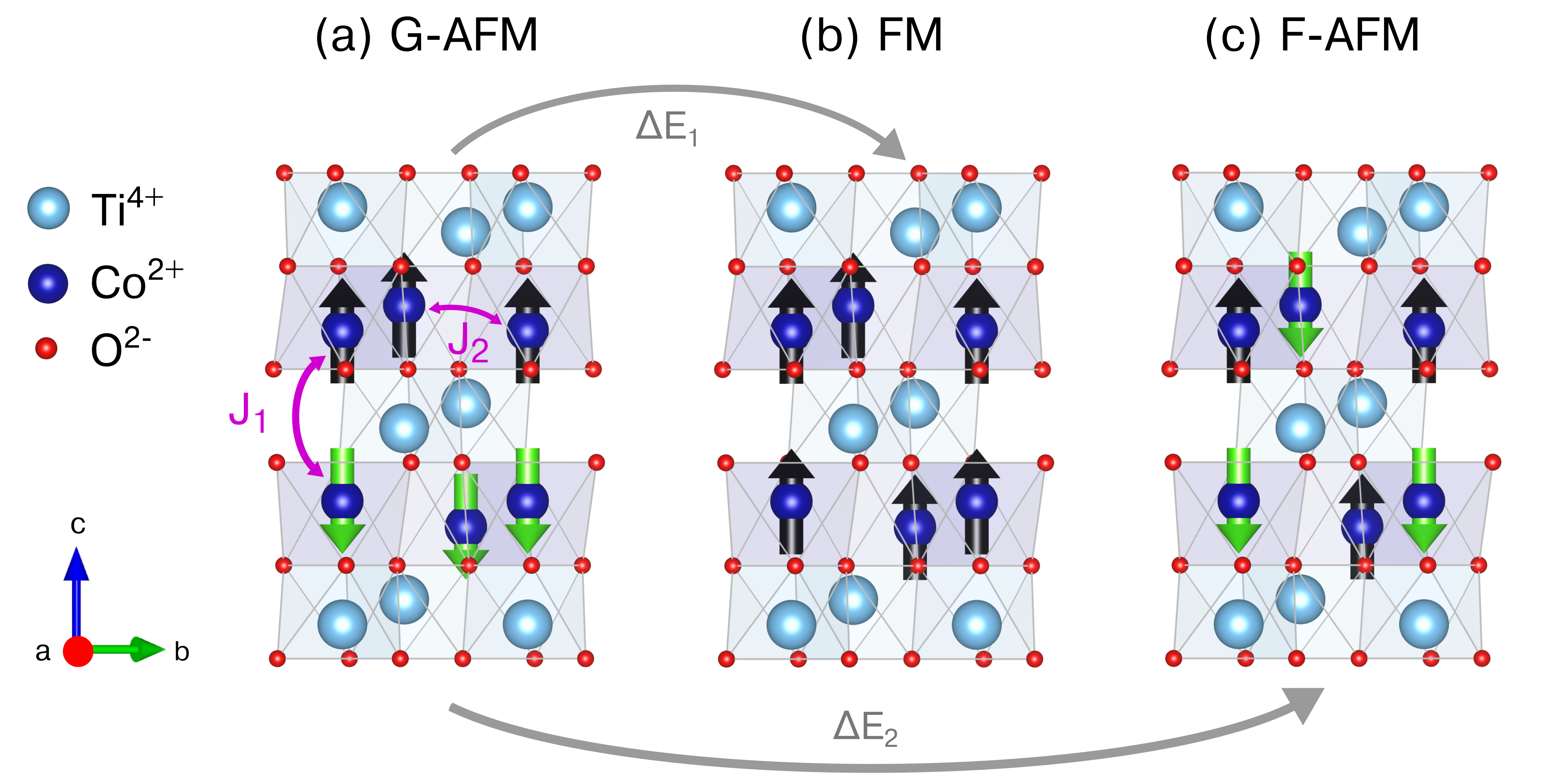}}
\caption{\label{fig:mag_structure}(a) ``G-type" antiferromagnetic, (b) ferromagnetic and (c) full-antiferromagnetic configurations of ilmenite CoTiO$_3$. $J_1$ and $J_2$ are the inter-layer and intra-layer magnetic couplings, respectively, and $\Delta E_1$ and $\Delta E_2$ denote the energetic differences between the configurations. In the (a) and (b) settings, the intra-layer coupling is ferromagnetic, with antiferromagnetic (a), or ferromagnetic (b), inter-layer coupling. In the configuration (c), both couplings are antiferromagnetic. }
\end{figure*} 

\subsection{Magnetic properties} 

\subsubsection{Spin configuration}

To analyze the magnetic structure of cobalt titanate, we perform total energy calculations for various magnetic configurations: the G-AFM structure, the ferromagnetic (FM) and ``full-antiferromagnetic" (F-AFM) structures shown in Fig.\ref{fig:mag_structure}. 
We find that for all considered ($U_{\text{Ti}}$,$U_{\text{Co}}$) pairs, the energy ordering of the three structures is the same: the G-AFM configuration is the ground-state of the system, followed by the ferromagnetic FM state, with the full-antiferromagnetic F-AFM structure presenting a considerable higher energy. We refer to the energetic difference between the G-AFM and the FM structures as $\Delta E_1$, and label the difference between the G-AFM and F-AFM states as $\Delta E_2$.

In the G-AFM state, all cobalt atoms have a local magnetic moment of $\pm |\mu_{Co}|$, where $|\mu_{Co}|$ ranges from 2.5 $\mu_{B}$ (GGA) to 2.8 $\mu_{B}$ ($U_{\text{Ti}}$=6, $U_{\text{Co}}$=5). This change in the local magnetic moment is also the cause of the localization effect due to the $U$ parameters, which concentrates the electronic density around the cobalt atoms as the $U$ parameters increase. The calculated magnetic moments are close to the expected S=3/2 value derived from Hund's rules, and the slight difference can be attributed to the fact that the local magnetization is numerically computed by integrating in the spherical region given by the Wigner-Seitz radius, which can lead to an underestimation of the measured magnetization. However, it should not be forgotten that due to the hybridization mentioned in the previous section, cobalt presents a non-negligible covalence that modifies the ionic Co$^{2+}$ picture.

\begin{figure}
    \centering
    \includegraphics[scale=0.46]{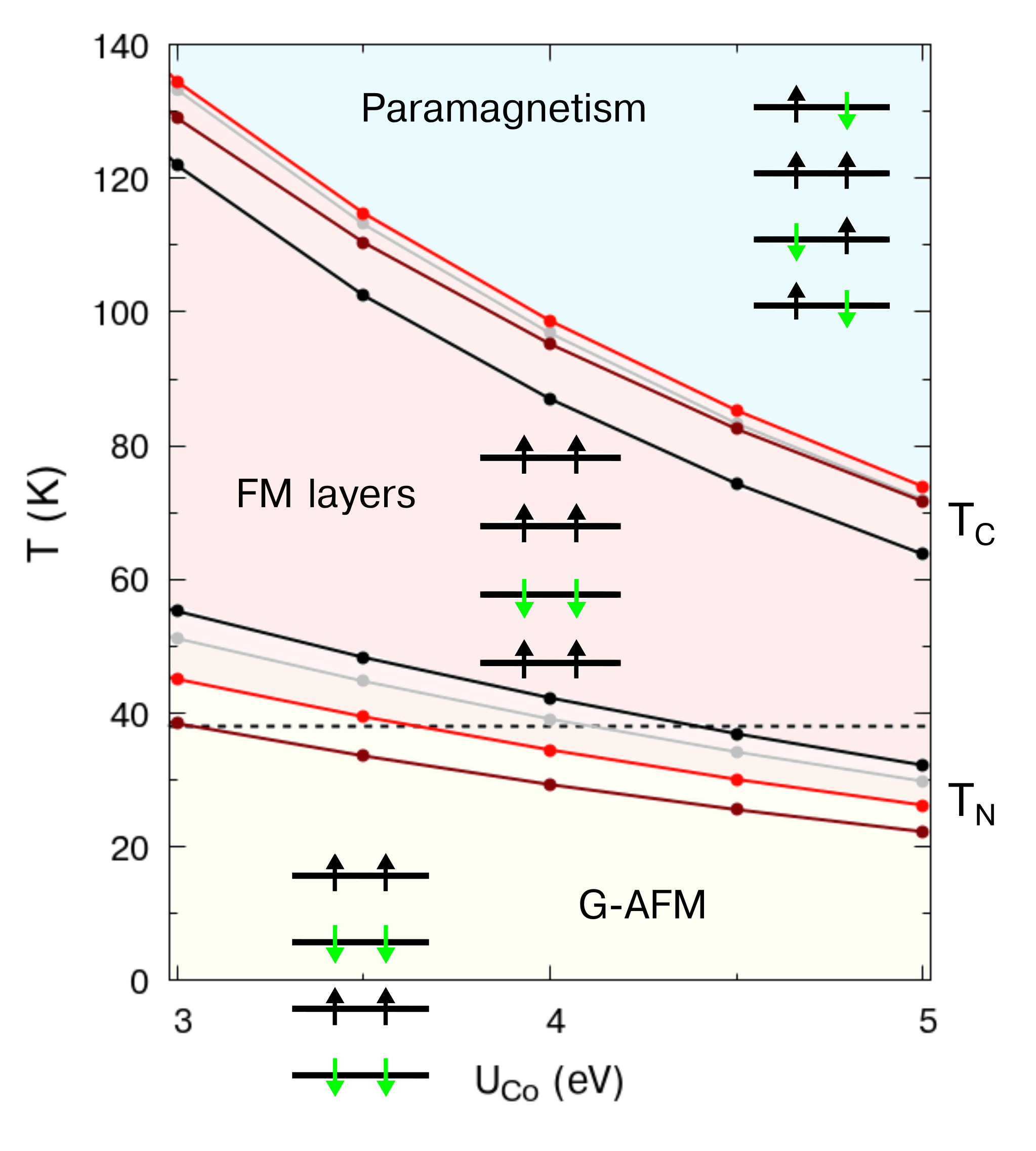}
    \caption{Magnetic phases of cobalt titanate with temperature. Energy differences $\Delta E_1$ and $\Delta E_2$ were converted to temperature units. T$_{\text{N}}$ is the N\'{e}el temperature, which is related to the breaking of the inter-layer antiferromagnetic ordering, and is shown along the experimental value of T$_{\text{N}}$=38K indicated by a dashed line. T$_{\text{C}}$ is related to the breaking of the intra-layer ferromagnetic ordering, and represents the starting point of paramagnetic behavior which exists for higher temperatures. 
     }  
    \label{fig:energy_diff}
\end{figure}

The previously defined energy differences $\Delta E_1$ and $\Delta E_2$ can be linked to the thermal energy needed to invert the spin ordering of their respective coupling, which causes a phase transition. A critical temperature can be associated with each of these transitions, e.g. in the form of $k_B T_i = \Delta E_i/N_{Co}$. The inter-layer superexchange $J_1$ and intra-layer direct exchange $J_2$ couplings can also be calculated from these energies (see Appendix D). We get approximate values of $J_1$ = 1.33 meV and $J_2$ = -1.25 meV in the $U_{\text{Co}}$=$U_{\text{Ti}}$=4.0 eV case. These values have not to be confused with the ones in Refs.\cite{DiracMagnons,DUBROVIN2020157633}, which are calculated for different model Hamiltonians and other DFT approaches. %\\

These computed critical temperatures are shown in Fig.\ref{fig:energy_diff}, where a phase diagram of the system behavior is presented. 
For temperatures lower than the N\'{e}el temperature (T$_N$), the system will exhibit the G-AFM state, which consists of ferromagnetic hexagonal $ab$ planes antiferromagnetically coupled along the $c$-axis. When the temperature ranges between T$_N$ and T$_C$, the antiferromagnetic inter-layer ordering will be broken, but the intra-layer ferromagnetic ordering will still be present. Lastly, T$_C$ indicates the beginning of the fully paramagnetic behavior, where the thermal energy overcomes the in-layer coupling, breaking the ferromagnetic ordering of the layers. Note that the Co ions in the paramagnetic state still present disordered local magnetizations, not being fully spin compensated.
These findings suggest that individual layers can be ferromagnetic in the T$_N <$ T $<$ T$_C$ range above the N\'{e}el temperature, an interesting result regarding applications that might merit further experimental work.

\subsubsection{Magnetic anisotropy} 

%MAGNETIC ANISOTROPY:  

\begin{figure}
    \centering
    \includegraphics[scale=0.4]{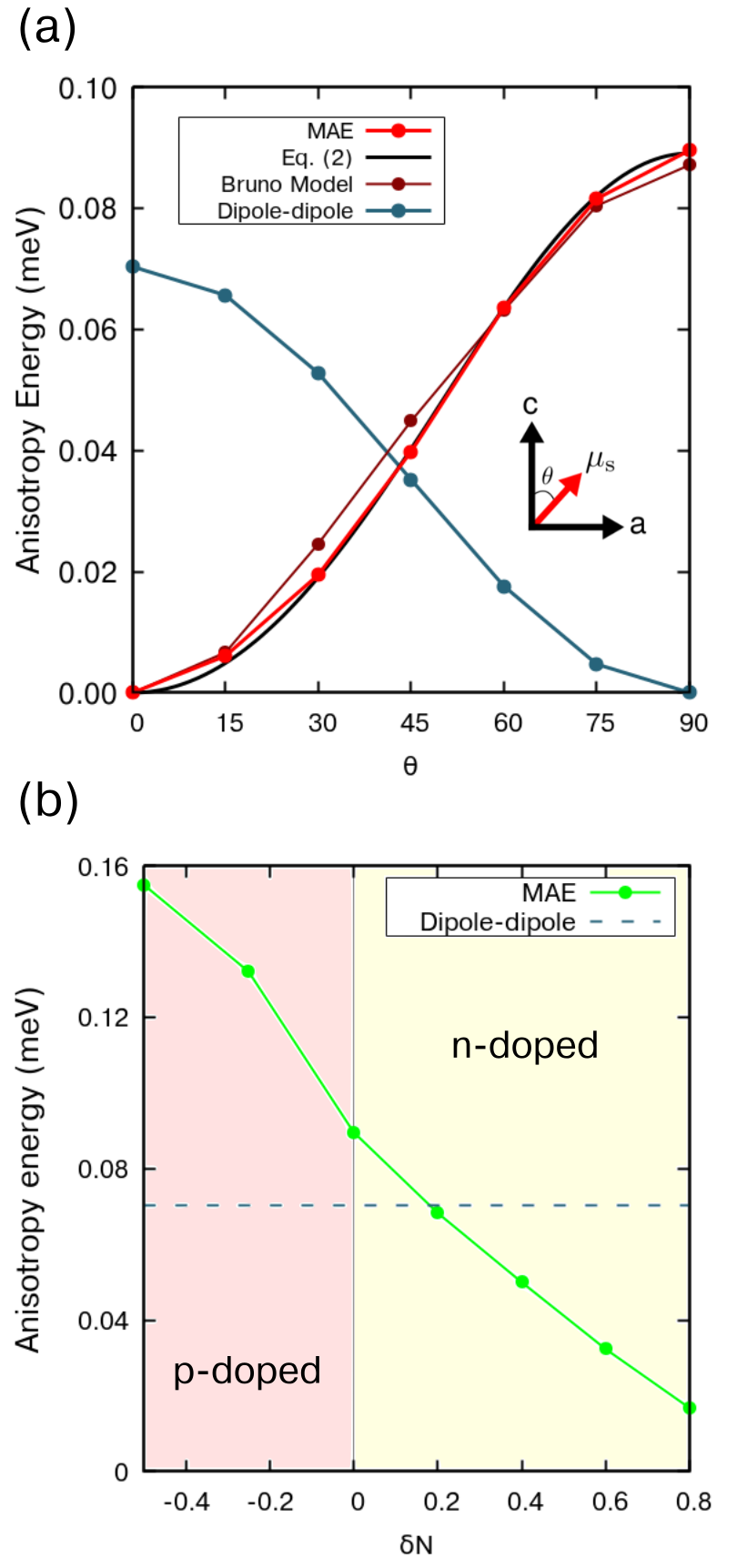}
    \caption{Magnetic anisotropy of bulk cobalt titanate. (a) Angular dependence of the anisotropy, where $\theta$ changes from the out-of-plane \textit{c}-axis to the hexagonal \textit{ab} plane. In red, the MAE is shown as calculated using DFT. In dark red, the Bruno model fitting obtained with the calculated $\mu_L$ orbital moments is shown; in black, the fitting to Eq.\eqref{eq:MAE_sin}; in blue, the magnetic anisotropy due to the dipole-dipole interaction is shown. 
    (b) Magnetic anisotropy at the \textit{ab} plane ($\theta = 90\degree$) with respect to the variation of the electron number in the unit cell ($\delta$N). 
    The dipole-dipole term is shown in dashed lines. For an electron excess of around 0.2, the MAE term becomes smaller than the dipole-dipole term. 
    %Even if it slightly changes due to the variation in the magnetic moment with $\delta N$, it is plotted as a constant, as its variation can hardly be distinguished by eyesight in the scale range shown in the figure. 
    }
    \label{fig:Anisotropy}
\end{figure}

%Intro: 
We next consider the magnetic anisotropy due to the ferromagnetic cobalt layers in CoTiO$_3$ bulk. 
In order to determine whether cobalt titanate presents an in-plane or out-of-plane magnetic anisotropy, we first perform total energy calculations including the spin-orbit term as implemented in VASP for a number of spin orientations with respect to the ferromagnetic cobalt layers. The magnetocrystalline anisotropy energy (MAE) is defined as the energetic difference between the lowest energy magnetic configuration and the configuration under analysis, and is given by MAE($\theta$) = $E(\theta$) - $E_{\text{z}}$. 
Here, $\theta$ is the polar angle in the hexagonal $ac$ (cartesian $xz$) plane. 
We found that the effect of the in-plane orientation was negligible, only varying the MAE in the order of $\mu$eV for different values of the azimuthal angle within the hexagonal $ab$ plane.
In the G-AFM setting of the primitive magnetic cell, we calculated the MAE in the GGA and GGA+U approaches, with $U_{\text{Ti}}$=3.9 eV and $U_{\text{Co}}$=4.5 eV. The MAE values using GGA are larger than those for the GGA+U cases because the GGA structure is slightly compressed. In fact, the role of the structural parameters seems key as the MAE for the experimental lattice is even larger. Some comments on the effect of U in the anisotropy can be found in Appendix F. We then focus on the MAE per atom in the GGA+U case, as shown in Fig.\ref{fig:Anisotropy}(a).

%Angle dependence: 
We observe that the magnetocrystalline anisotropy is minimum in the out-of-plane hexagonal $c$-axis, and increases as spins align with the hexagonal $ab$ plane. This tendency is observed in both the GGA and GGA+U approaches, suggesting the easy-axis character of the hexagonal $c$-axis. The MAE was also calculated for the ferromagnetic configuration and found to be out-of-plane. This finding confirms that the ab layers have a strong out-of-plane character.
In order to understand the angular dependence of the MAE, we fit our results to the expresion 
\begin{equation}
    \text{MAE}(\theta) = K_{1} \sin^2{(\theta)} + K_2 \sin^4{(\theta)}, 
    \label{eq:MAE_sin}
\end{equation}
where $K_1$ and $K_2$ are the magnetocrystalline anisotropy constants\cite{MAE-Chikazumi}. Using the total energies per unit cell, our fitting yields values of $K_1$ = 0.29 (0.52) meV and $K_2$ = 0.068 (0.025) meV for GGA+U (GGA) cases. The  $K_1$ value is much larger than the $K_2$ one, but not negligible for GGA+U. This  indicates the strong uniaxial character of the anisotropy.  \\ 

%Element dependence: 
The element dependence of the anisotropy can also be analyzed by fitting the MAE to the Bruno model\cite{Bruno-PhysRevB.39.865} given by 
\begin{equation}
    \text{MAE}(\theta) = -\frac{\xi}{4\mu_B} (\mu_L^{GS} - \mu_L(\theta)) > 0,  
    \label{eq:Bruno}
\end{equation} 
where  $\xi\simeq$ 50 meV is the spin-orbit constant,  and $\mu_L^{GS}$ and $\mu_L(\theta)$ are the orbital magnetic moments of cobalt atoms in the ground-state configuration and in the axis under analysis, respectively. 
Our fitting to Eq.(\ref{eq:Bruno}) yielded a spin-orbit constant of $\xi \simeq$ 48 meV ($\xi \simeq$ 60 meV)  in the GGA+U (GGA) approach, close to the aforementioned value.
These MAE values calculated from the orbital magnetic momenta $\mu_L$ using the Bruno model are in great agreement with the directly calculated DFT+U values. This agreement suggests that the MAE could be directly correlated to the angular dependence of the density around cobalt ions in CoTiO$_3$. 
In the GGA+U approach, we get values of $\mu_L$ between 0.16 and 0.19 $\mu_B$, in good agreement with the only-GGA values in Ref.\cite{Das}. By being non-negligible, these $\mu_L$ values are pointing to the relevance of spin-orbit coupling in these cobaltates. The $\mu_L$ values are noncollinear with $\mu_S$ ones when the field is not exactly aligned with the easy axis or the hard plane (see Appendix G).

%Discussion in/out-of-plane : 
Previous reports point to an in-plane anisotropy in cobalt titanate \cite{Newnham,Elliot,DiracMagnons,WATANABE1980}, which is in contrast to our calculations. In order to understand this discrepancy, we also calculated the anisotropy due to the magnetic dipole-dipole interaction\cite{Bruno-dipolar-inbook}. This interaction is given by the term 
\begin{equation}
    H_{d-d} = -\sum_{i\ne j}{\frac{\mu_0}{4\pi |\textbf{r}_{ij}|^3} \biggl(3 (\textbf{m}_i \cdot \hat{\textbf{r}}_{ij})(\textbf{m}_j \cdot \hat{\textbf{r}}_{ij}) - \textbf{m}_i\cdot \textbf{m}_j  \biggr)}, 
\end{equation}
where \textbf{m}$_i$ and \textbf{m}$_j$ are the local magnetic moments around the interacting cobalt ions pairs, and \textbf{r}$_{ij}$ is the vector joining the two cobalt atoms. We computed this term from the atomic positions and local magnetic moments derived from the DFT calculations in which the spin-orbit interaction was included. Our results for the GGA+U structure are shown along the MAE in Fig. \ref{fig:Anisotropy}(a).  

In contrast to the spin-orbit term, the dipole term favors in-plane spin orientation, and competes with the MAE term in magnitude. Nevertheless, the total magnetic anisotropy still favors an out-of-plane orientation in our calculations. This effect presumably increases with growing temperature, 
as the dipole-dipole term (approximately $\propto M^2(T)$) decays faster than the MAE term ($\propto M(T)$) with the spontaneous magnetization\cite{Kittel_magnetization}. This could lead to potential out-of-plane ferromagnetic layers in the T$_{\text{N}} <$ T $<$T$_{\text{C}}$ temperature range.

To reconcile our results with experiments, we analyze the effect of doping in the system, see Fig. \ref{fig:Anisotropy}(b). This is performed by the addition and the substraction of electrons in the unit cell. Including defects in this compound explicitly implies a different set of calculations beyond the scope of the actual paper. We find that removing electrons (p-doping) leads to an increase of the MAE, while adding electrons (n-doping) lowers the MAE even past the dipole-dipole term. This later mechanism could be a consequence of the presence of Ti atoms at some cobalt sites in the sample, as suggested in the experimental literature\cite{Shirane, Newnham}. Our results indicate that adding 0.2 electrons (which roughly corresponds to 2.5\% of cobalt sites being occupated by titanium) could be enough to turn the out-of-plane anisotropy to an in-plane anisotropy, consistent with experiment.
We further remark that the effect of mesoscopic domains, suggested in the literature\cite{DiracMagnons,Elliot}, may result in domains with in-plane anisotropy. However, it should be noted that domains with an out-of-plane component could also lead to the compensation of the MAE, yielding an in-plane anisotropy, as already shown in magnetic alloys \cite{enkovaara2002magnetic,enkovaara2002structural}. 

In summary, we find that crystalline bulk CoTiO$_3$ presents a strong out-of-plane magnetocrystalline anisotropy, due to the spin-orbit coupling of cobalt atoms. The value is larger in magnitude to that of pure hcp cobalt\cite{daalderop1988magnetic,daalderop1990firstprinciples}, a fact that is interesting because cobalt can be seen in this compound as a Co$^{2+}$ ion instead of being metallic.  Furthermore, the dipole-dipole interaction is also estimated to be significant in this material due to cobalt ferromagnetic coupling in layers. Summing the two contributions, we observed that the presence of cobalt-titanium anti-site disorder could be responsible of the experimentally observed in-plane anisotropy of the bulk CoTiO$_3$.

%%%%%%%%%%%%%%%%%%%%%%%%%%% %%%%%%%%%%%%%%%%%%%%%%%%%%% %%%%%%%%%%%%%%%%%%%%%%%%%%%  
%%%%%%%%%%%%%%%%%%%%%%%%%%%         CONCLUSIONS         %%%%%%%%%%%%%%%%%%%%%%%%%%% 
%%%%%%%%%%%%%%%%%%%%%%%%%%% %%%%%%%%%%%%%%%%%%%%%%%%%%% %%%%%%%%%%%%%%%%%%%%%%%%%%% 

\section{Conclusions}
\label{sec:conclusions}
In this paper we analyzed the structural, electronic and magnetic properties of ilmenite CoTiO$_3$ in the DFT+U framework. We observed that while the addition of the U correction terms slightly expands the unit cell of the system, it greatly improves the description of the electronic properties by partially correcting the electron delocalization, and thus enhancing the semiconducting character of the system. 

Regarding the magnetism of cobalt titanate, we found that the G-AFM structure is the ground state of the system, and that there are two critical temperatures which correspond to the transition between the G-AFM and ferromagnetic-layered structure, and to the beginning of the paramagnetic phase. The existence of ferromagnetic planes at temperatures above T$_{\text{N}}$, could potentially lead to interesting magnetic applications.

Our calculations including spin-orbit coupling indicate that the anisotropy would be out-of-plane, a finding in contrast with experiments.  However, we found that the presence of 0.2 electrons in the unit cell (which roughly corresponds to 2.5\% of cobalt sites being occupated by titanium) could be enough to turn the out-of-plane anisotropy to an in-plane anisotropy, consistent with experiment. We believe that further experimental studies, such as high pressure experiments, could further deepen our understanding of the magnetic anisotropy in this material. On the theoretical front, slab and single-layer calculations seem of great interest for future investigations of intriguing thin-film systems.

\section{Acknowledgements}
We gratefully acknowledge primary funding from the National Science Foundation through the Center for Dynamics and Control of Materials:an NSF MRSEC under Cooperative Agreement No. DMR-1720595, with additional support from NSF DMR-1949701 and NSF DMR-2114825.  This work was performed in part at the Aspen Center for Physics, which is supported by National Science Foundation grant PHY-1607611. M. R-V. was supported by LANL LDRD Program and by the U.S. Department of Energy, Office of Science, Basic Energy Sciences, Materials Sciences and Engineering Division,
Condensed Matter Theory Program.

This work has been supported by the Spanish Ministry of Science and Innovation with PID2019-105488GB-I00 and PCI2019-103657. 
We acknowledge financial support by the European Commission from the NRG-STORAGE project (GA 870114). 
The Basque Government supported this work through Project No. IT-1246-19. M.A was supported by the Spanish Ministry of Science and Innovation through the FPI PhD Fellowship BES-2017-079677.

%%%%%%%%%%%%%%%%%%%%%%%%%%%%%%%%%%%%%%%%%%%%%%%%%%%%%%%%%%%%%
%%%%%%%%%%%%%%%%%%%%%%%%%%%APPENDIX%%%%%%%%%%%%%%%%%%%%%%%%%%
%%%%%%%%%%%%%%%%%%%%%%%%%%%%%%%%%%%%%%%%%%%%%%%%%%%%%%%%%%%%% 

\appendix

\section{Tests with an all-electron method}
In order to check the validity of the chosen number of valence electrons to be included per element, we compare the VASP calculation with a more precise all-electron calculation performed with the Elk code\cite{elk}. We find that both band structures are in great qualitative agreement, which confirms the validity of the chosen number of valence electrons per element in our calculations. 

\begin{figure}
    \centering
    \includegraphics[scale=0.3]{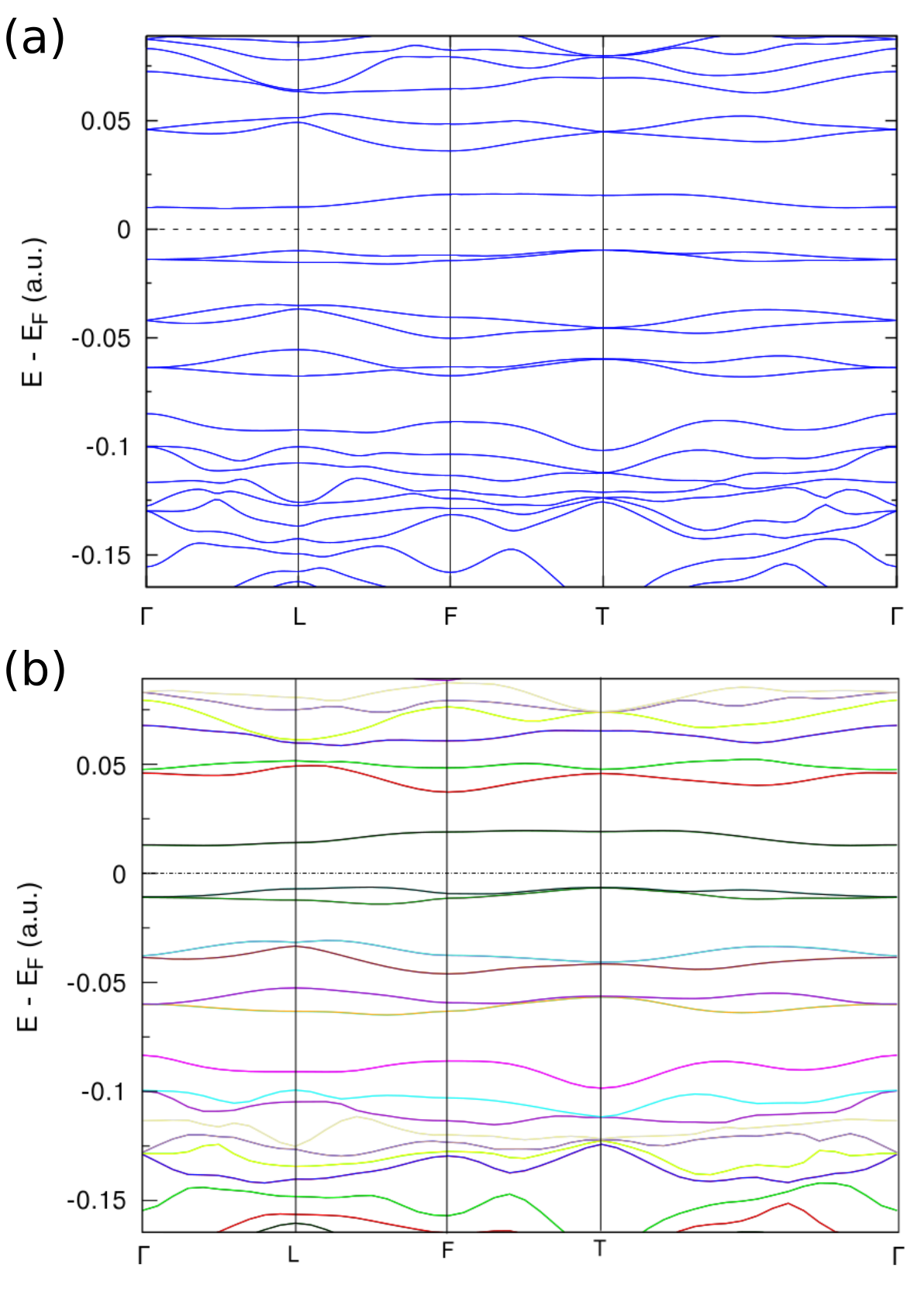}
    \caption{F-AFM band structure in the chemical primitive cell, in the GGA (panel (a)) and all-electron (panel (b)) approaches.}
    \label{fig:dft-elk}
\end{figure}

\section{Phonon calculations}
Lattice-dynamics calculations were performed using the supercell finite-displacement method implemented in the Phonopy software package\cite{Phonons}, with VASP used as the 2nd order force-constant calculator. Calculations of the phonon supercell size were carried out on 2×2×2 expansions of the primitive-cell. For the DFT force calculations, we employ spin-collinear formalism with an energy cut-off of 700 eV, a 6 x 6 x 6 k-point Monkhorst pack mesh and the DFT-D3 Grimme\cite{Grimme} van der Waals correction method. The U parameters where chosen to be: U$_{\text{Ti}}$ = 3.9 eV and U$_{\text{Co}}$ = 4.5 eV. 

\begin{figure}
    \centering
    \includegraphics[scale=0.375]{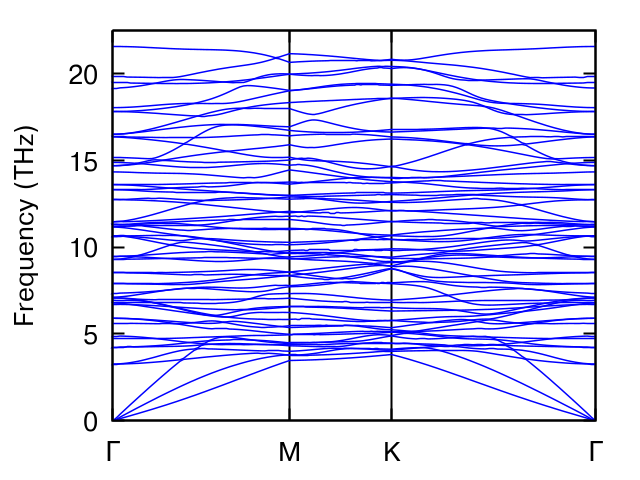}
    \caption{Phonon dispersion of the primitive magnetic cell in the G-AFM structure.}
    \label{fig:phonons}
\end{figure}

\section{Electronic structure using GGA+U approach}
As mentioned in the main text, the effect of the cobalt U parameter is to split the cobalt bands localized around the Fermi energy, gradually increasing the band gap and enhancing the hybridization with titanium and oxygen bands. 

\begin{figure}[h]
    \centering
    \includegraphics[scale=0.4]{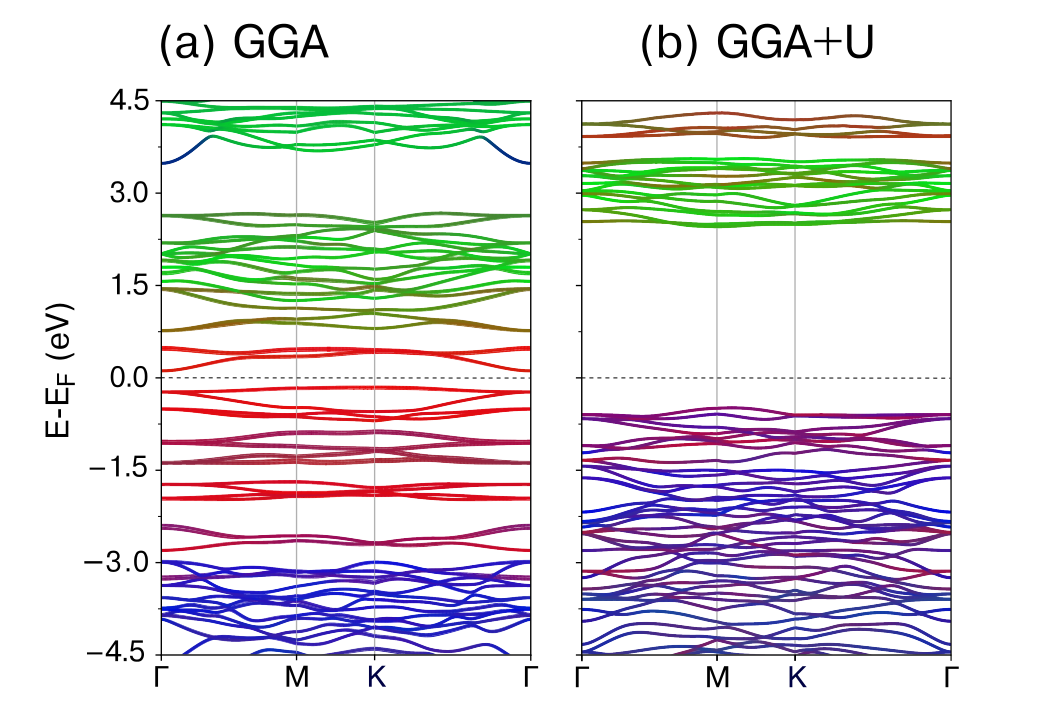}
    \caption{(a) GGA and (b) GGA+U element-projected band structures of cobalt titanate. U parameters U$_{\text{Co}}$ = U$_{\text{Ti}}$ = 4.0 eV were chosen. Color indicates the fractional character of each element in the bands, with cobalt given in red, titanium in green and oxygen in blue. These band structures were calculated using the sumo software\cite{Sumo}.}
    \label{fig:gga-bands}
\end{figure}

\section{Calculation of the magnetic coupling constants}
In the magnetic configurations under analysis, the Heisenberg Hamiltonian 
\begin{equation}
    \text{H} = \sum_{ij} J_{ij} \textbf{S}_i \cdot \textbf{S}_j,  
    \label{eq:Heisenberg}
\end{equation}
yields the following energies per primitive magnetic cell: 
\begin{eqnarray}
    \text{E}_{\text{G-AFM}} &= (- 2J_1 + 6J_2)\Tilde{S}^2, \\ 
    \text{E}_\text{{FM}}    &= (2J_1 + 6J_2)\Tilde{S}^2, \\ 
    \text{E}_\text{{F-AFM}} &= (-2J_1 - 6J_2 )\Tilde{S}^2.
\end{eqnarray}
Here, $\Tilde{S}$ is the pseudospin 3/2, and $J_1$ and $J_2$ are the inter-layer and intra-layer magnetic couplings (given in meV). From the energy differences $\Delta \text{E}_1 = \text{E}_{\text{FM}} - \text{E}_{\text{G-AFM}}$ and $\Delta \text{E}_2 = \text{E}_{\text{F-AFM}} - \text{E}_{\text{G-AFM}}$, we get the following expressions for the couplings: 
\begin{eqnarray}
    J_1 = \frac{\Delta \text{E}_1}{4\Tilde{S}^2}, \\ 
    J_2 = -\frac{\Delta \text{E}_2}{12\Tilde{S}^2}.
\end{eqnarray}

\section{Convergence of the MAE}
\begin{figure}
    \centering
    \includegraphics[scale=0.325]{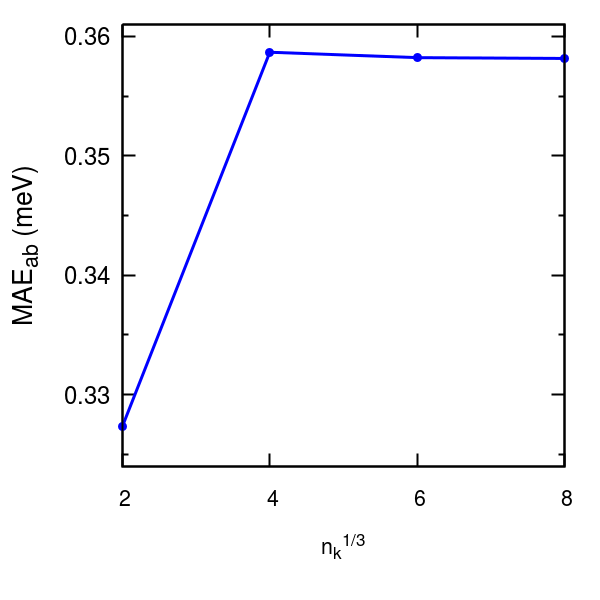}
    \caption{MAE of the hexagonal \textit{ab} plane against the Brillouin Zone sampling per unit cell. Regular $\Gamma$-point centered $n_{k}\times n_{k}\times n_{k}$ Monkhorst-Pack grids were used. }
    \label{fig:MAE-k}
\end{figure}

The MAE is a small magnitude, in our case in the order of $10^{-4}$ eV. To ensure that our results are numerically correct, we calculate the MAE amplitude with respect to the Brillouin Zone sampling $n_{k}$ (Fig. \ref{fig:MAE-k}). We show that the MAE has a fast convergence in this system, and that the 8x8x8 Monkhorst-Pack grid used in our calculations gives a well converged anisotropy energy.

\section{Effect of U in the magnetocrystalline anisotropy energy}
\begin{figure}
    \centering
    \includegraphics[scale=0.325]{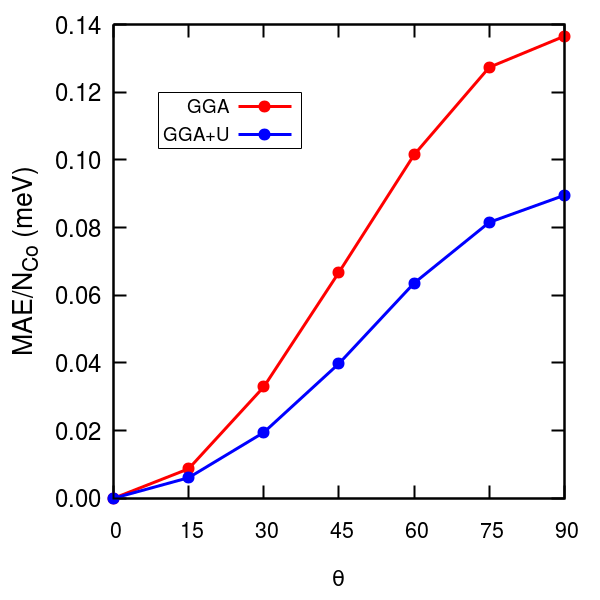}
    \caption{Comparison between the MAE in the GGA and GGA+U approaches. The MAE values are given per cobalt atom.}
    \label{fig:MAE-GGA}
\end{figure}

As commented in the main text, we find that the MAE values obtained in the GGA approach are larger than those using GGA+U, as shown in Fig. \ref{fig:MAE-GGA}.
The structural expansion induced by the U parameter plays a leading role in this trend, as the GGA structure is considerably closer to the experimental cell. The GGA+U anisotropy with the experimental lattice parameters is slightly larger  (MAE$_{\textit{ab}}/\text{N}_{\textit{Co}} \simeq $ 0.145 meV), a fact that points to the structural expansion as the main responsible for a decreasing MAE. 

\section{Orbital moment and non-collinearity}
We show the calculated orbital magnetic moment values $\mu_{\text{L}}$ in the GGA+U approach, as well as the angular difference $\Delta\theta$ between the spin and orbital magnetic moments that arises when the spin-orbit coupling is included. Note the overall non-collinearity between spin and orbital moments unless the $\theta$ values are just $\theta = 0, \pi/2$ and $\pi$.
\begin{figure}[h]
    \centering
    \includegraphics[scale=0.4]{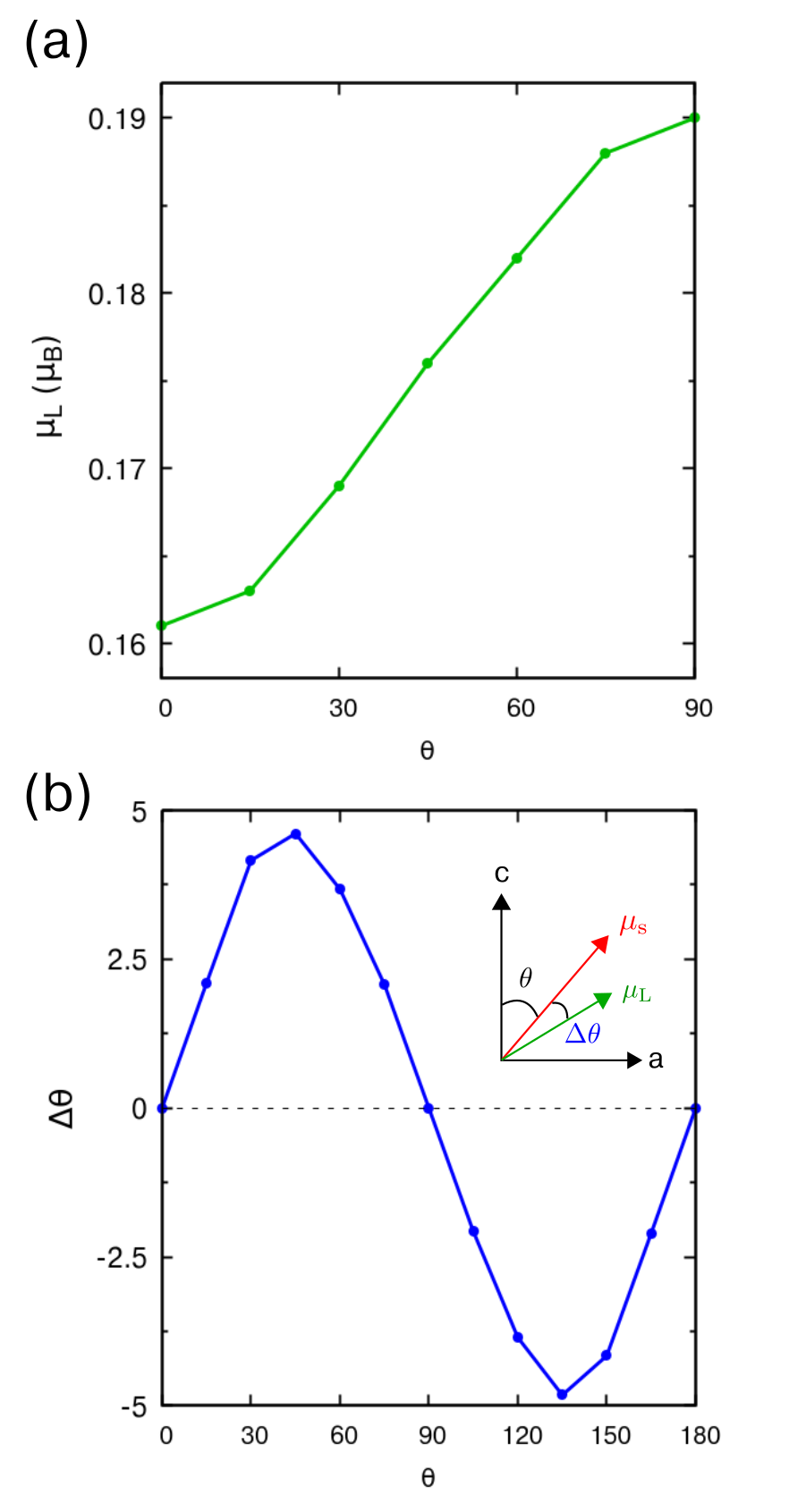}
    \caption{(a) Orbital moment $\mu_{\text{L}}$ of cobalt titanate calculated in the GGA+U approximation. (b) Angular difference $\Delta \theta$ between the spin and orbital magnetic moments.}
    \label{fig:non-collinearity}
\end{figure} 

\section{Charge density with doping}
\begin{figure}
    \centering
    \includegraphics[scale=0.15]{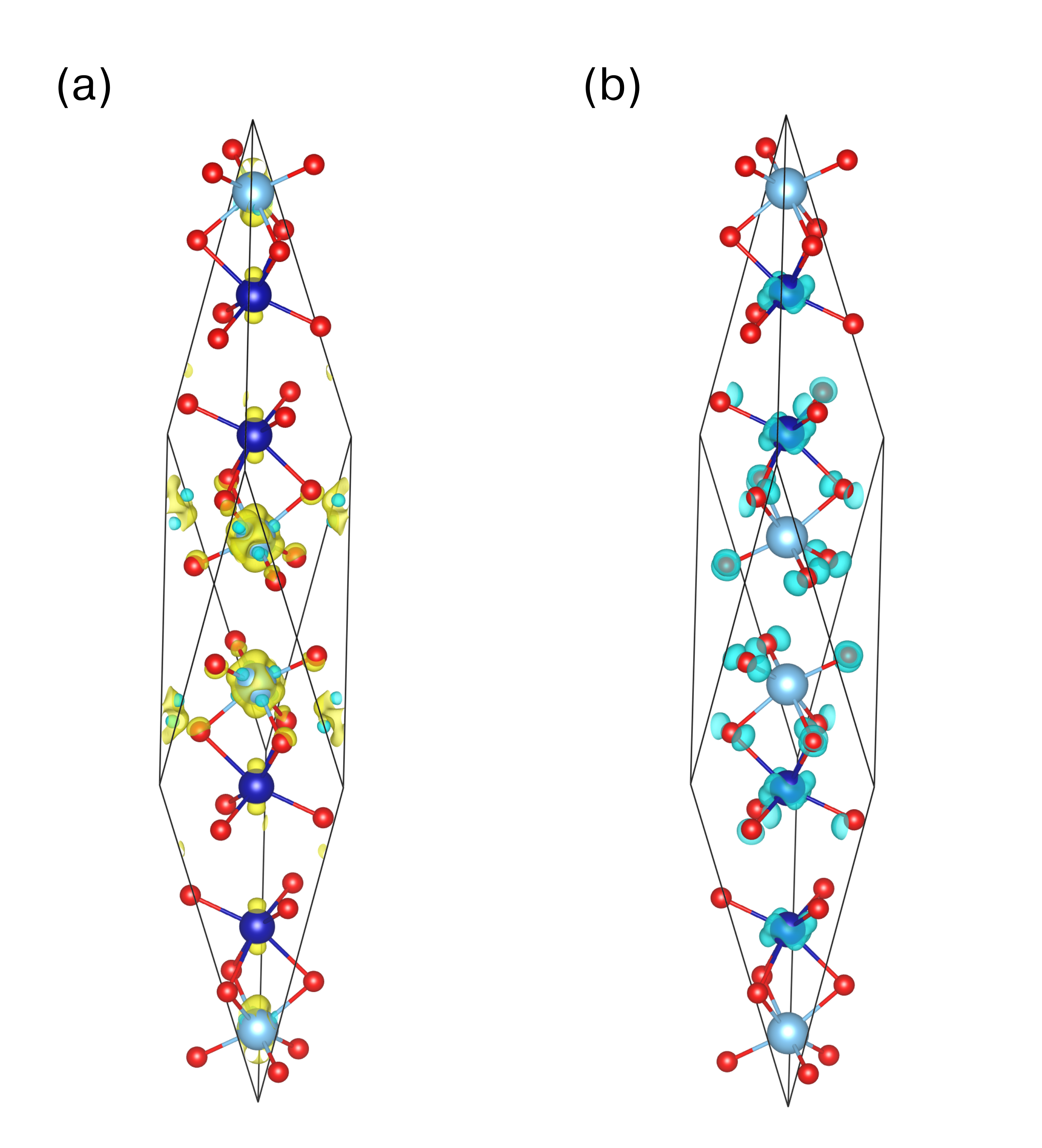}
    \caption{Charge density difference for the (a) n-doped and (b) p-doped cobalt titanate with respect to the non doped case. The density is plotted in units of $e/{a_0}^3$, where $e$ is the electron charge and $a_0$ is the Bohr radius. 
    The isosurface level is set to 0.00161 $e/{a_0}^3$ in panel (a) and to 0.00118 $e/{a_0}^3$ in panel (b). Yellow and cyan denote excess and deficit charge density difference, respectively. Charge densities were plotted using the VESTA software\cite{VESTA}. }
    \label{fig:CHG} 
\end{figure}

The addition (substraction) of electrons in the unit cell creates an excess (deficit) of charge. The charge density differences between the non doped CoTiO$_3$ and the n-doped (panel (a)) and p-doped (panel (b)) structures are shown in Fig.\ref{fig:CHG}. 
On the one hand, electrons localize around titanium atoms and the d$_{z}^2$ orbitals of cobalt atoms. On the other hand, holes are localized around the rest of d orbitals in cobalt and the p orbitals in oxygen atoms.   
This trend is in good agreement with the electronic band structure of CoTiO$_3$ shown in Fig. \ref{fig:bands_evolution}, where the bottom of the conduction band consists of titanium and cobalt d$_{z}^2$ orbitals, while the top of the valence band is a mixture of the rest of the d orbitals of cobalt and the p orbitals of oxygen. 

\bibliography{apssamp}% Produces the bibliography via BibTeX.

\end{document}